\documentclass{sf2a-conf2014}
\usepackage{graphicx}
\usepackage{hyperref}
\usepackage[]{natbib}  
\usepackage{epstopdf}
\usepackage{sidecap}

\def\BibTeX{{\rm B\kern-.05em{\sc i\kern-.025em b}\kern-.08em
    T\kern-.1667em\lower.7ex\hbox{E}\kern-.125emX}}
\bibpunct{(}{)}{;}{a}{}{,}  

\begin{document}

\TitreGlobal{SF2A 2014}


\title{Thermal and radiative AGN feedback :\\ weak impact on star formation  in high-redshift disk galaxy simulations}

\runningtitle{AGN feedback : weak impact on star formation}

\author{O. Roos$^1$}
\author{S. Juneau$^1$}
\author{F. Bournaud$^1$}
\author{J. M. Gabor}\address{CEA-Saclay, 91190 Gif-sur-Yvette, France}




\setcounter{page}{237}


\maketitle


\begin{abstract}
Active Galactic Nuclei (AGNs) release huge amounts of energy in their host galaxies, which, if the coupling is sufficient, can affect the interstellar medium (ISM). We use a high-resolution simulation ($\sim6$ pc) of a z $\sim2$ star-forming galaxy hosting an AGN, to study this not yet well-understood coupling. In addition to the often considered small-scale thermal energy deposition by the AGN, which is implemented in the simulation, we model long-range photo-ionizing AGN radiation in post-processing, and quantify the impact of AGN feedback on the ability of the gas to form stars. Surprisingly, even though the AGN generates powerful outflows, the impact of AGN heating and photo-ionization on instantaneous star formation is weak~: the star formation rate decreases by a few percent at most, even in a quasar regime ($L_{bol}=10^{46.5}$~erg~s$^{-1}$). Furthermore, the reservoirs of atomic gas that are expected to form stars on a 100~-~200~Myrs time scale are also marginally affected. Therefore, while the AGN-driven outflows can remove substantial amounts of gas in the long term, the impact of AGN feedback on the star formation efficiency in the ISM of high-redshift galaxies is marginal, even when long-range radiative effects are taken into account.
\end{abstract}



\section{Introduction}

Both observations and simulations give contradicting clues about the role of Active Galactic Nuclei (AGNs)  in galaxy evolution, and especially about their impact on star formation (SF). As resolution improved, simulations started to show that AGNs can both quench galaxies by expelling all their gas through outflows and preventing it from falling back \citep[e.g.][]{DiMatteo2005}, $and$ trigger SF through jet-induced shock waves \citep{Gaibler2012,Dugan2014}, and jet- or wind-induced ram pressure \citep{Silk2013}. Observationnally, evidence favoring both mechanisms can be found (e.g. \citet{Feain2007,Elbaz2009} for jet-induced SF ; \citet{Schawinski2007} for negative AGN feedback). Lately, studies also began to show that AGN outflows could have no impact on the global star formation rate (SFR) of their host \citep{Gabor2014}, and that, averaged over an extended period of time, all star-forming galaxies (SFGs)  host transient active episodes \citep{Hickox2014}, suggesting that the impact of AGN on SF could strongly depend on the time scale considered. 

AGN feedback can be seen as twofold in SFGs\footnote{We do not consider radio mode feedback (jets), which mainly affects early-type galaxies by maintaining their halo hot.}~: a thermal part, heating and pushing the gas away from the galactic center, creating outflows ; and a radiative part, ionizing gas and creating radiative pressure in the ISM. Until now, it is computationally difficult to run a simulation with both high resolution and a complete treatment of the radiative transfer (RT). However, many authors have shown that accounting for AGN photo-ionization could significantly change the properties of the ISM \citep{Maloney1999,Proga2014} and therefore the SFR of the galaxy. Different approaches can be used to bypass this problem: several teams implemented simplified RT or ionization computation, directly in their simulations, with a lower resolution ISM \citep[e.g.][]{Rosdahl2013,Vogelsberger2014c}. We treat RT in post-processing, in order to keep a high resolution allowing us to probe the $instantaneous$ effects of AGN radiation on a multi-phase ISM, with well-resolved giant molecular clouds (GMCs) and report the results presented in \citet{Roos2014}.

\section{Distribution of ionized gas and instantaneous SFR reduction}

\begin{figure}[ht!]
 \centering   
  \includegraphics[width=\textwidth,clip]{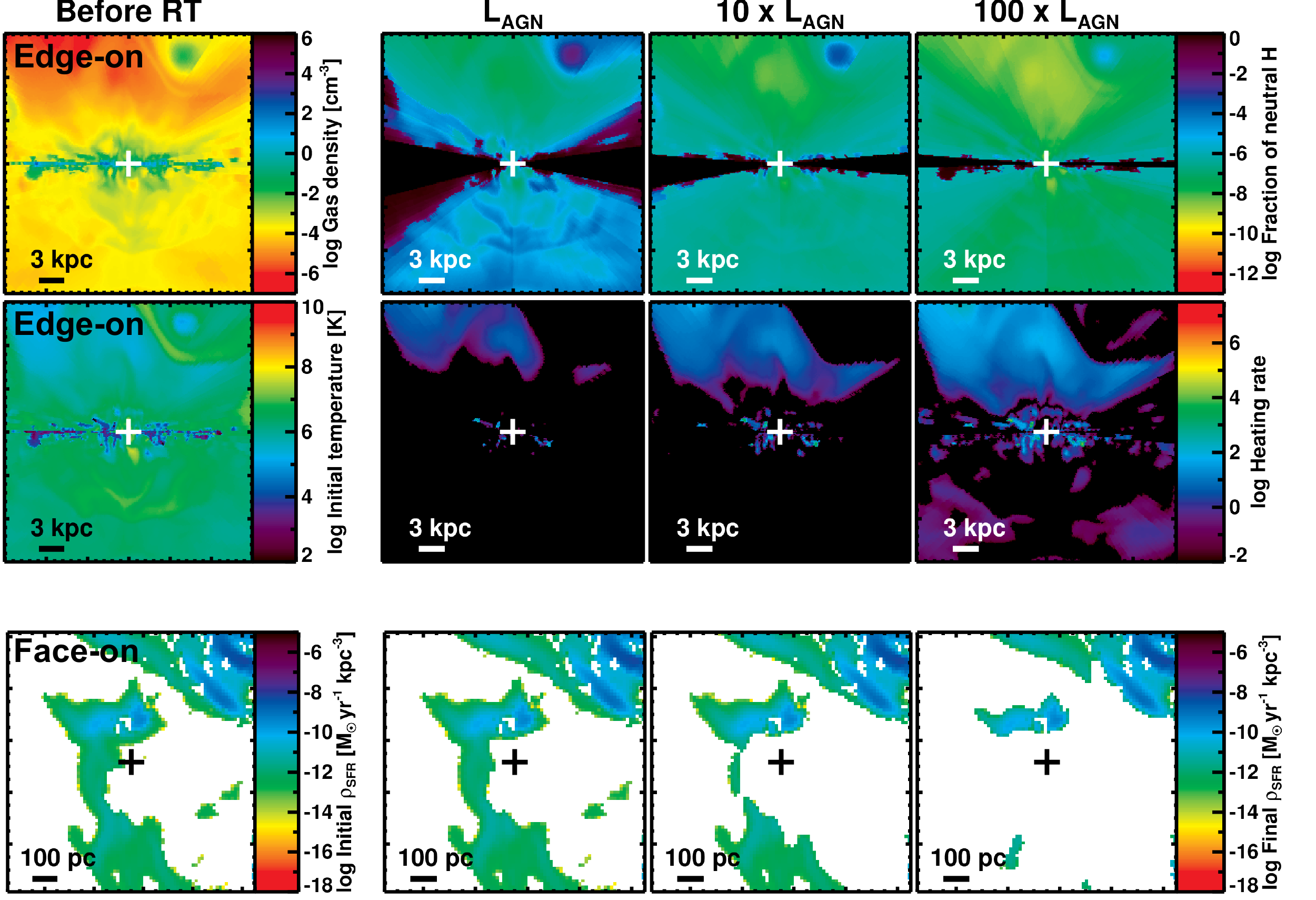}   
  \caption{Maps of a thin slice of the RT-processed galaxy for one representative snapshot. The two upper rows show the galactic disk edge-on at large scale. The bottom row displays a zoom on the AGN (`+' symbol), with a face-on view of the galactic disk. Left column shows the gas density, temperature and density of SFR before RT. The right columns show the effect of AGN ionization for the three AGN regimes, with the ionization fraction of hydrogen (top), the heating rate (middle) and the density of SFR (bottom). Even if most of the halo is ionized/heated by the AGN in all regimes, only the diffuse star-forming regions at the center of the galaxy ($< 1$ kpc) are prevented from forming stars due to AGN photo-ionization, and the bulk of the star-forming gas remains unaffected.}
  \label{roos:fig1}
\end{figure}

We use 6 snapshots of a high-resolution ($\sim6$ pc) simulation representing an isolated clumpy SFG at redshift~2 \citep{Gabor2013b}, including standard thermal AGN feedback \citep{Booth2009}. We post-process them with the RT code Cloudy \citep[last described by][]{Ferland2013}, as introduced by \citet{Roos2014}. This allows us to compute the combined effect of AGN photo-ionization and thermal feedback on the ionization state and temperature of the gas, and therefore on the SFR of the galaxy. However, the RT study is instantaneous and we cannot directly probe coupling to long-term gas dynamics or AGN variability. In the RT-process, the AGN is considered as the only ionizing source, and gas is considered initially neutral \citep[see][arXiv:1405.7971 for further details]{Roos2014}. Three AGN luminosity regimes were studied:
\begin{itemize}
\item a typical AGN, with $L_{bol}=10^{44.5}$~erg~s$^{-1}$, present in $\sim30$ \% of typical SFGs with masses between $10^{10}$ and $10^{11}$ M$_\odot$ \citep{Mullaney2012b,Juneau2013},
\item a strong AGN, with $L_{bol}=10^{45.5}$~erg~s$^{-1}$, present in $\sim3$ \% of typical SFGs in the same mass range,
\item a typical QSO, with $L_{bol}=10^{46.5}$~erg~s$^{-1}$, which is rare in typical SFGs in the same mass range.
\end{itemize}•

Figure \ref{roos:fig1} displays maps of a thin slice of the simulated galaxy centered on the AGN, before and after the RT process. The effect of AGN photo-ionization and heating is clearly visible at large scale in the halo (top and middle rows), and the amount of heated/ionized gas increases with AGN luminosity. The propagation of  AGN radiation through the ISM highly depends on the distribution of gas into clumps. Indeed, the presence of holes between the dense star-forming clumps of the ISM allows QSO radiation to reach a radius of up to $\sim8$ kpc in the galactic disk, while dense clumps are able to screen it at very small scale length.
 However, the impact of AGN feedback on the SFR of the galaxy remains small at all AGN luminosities, since only diffuse star-forming regions in the center of the galaxy (up to 1 kpc radius in the QSO regime) are prevented from forming stars.

\begin{figure}[ht!]
 \centering
 \includegraphics[width=0.58\textwidth,clip]{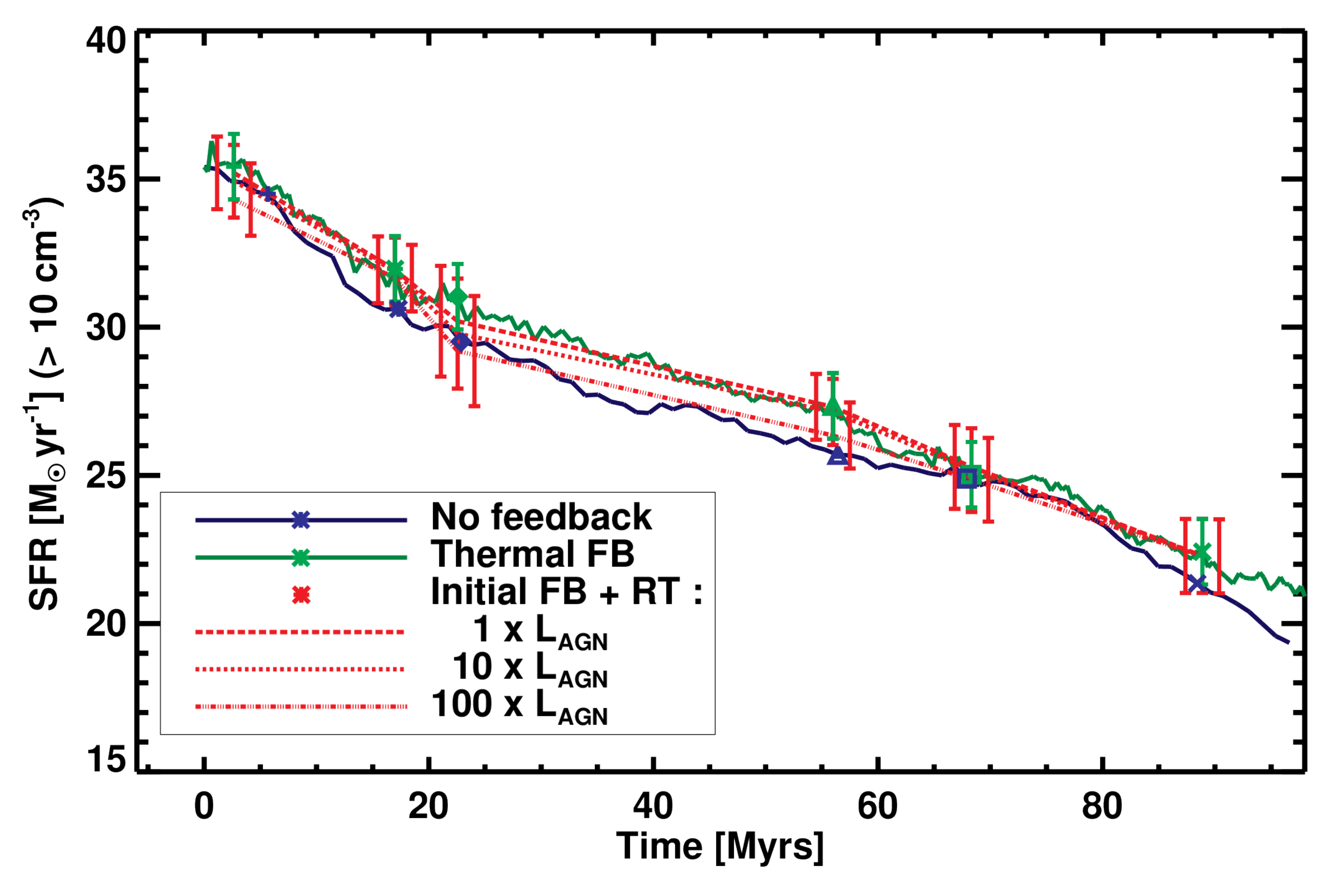}     
  \includegraphics[width=0.39\textwidth,clip]{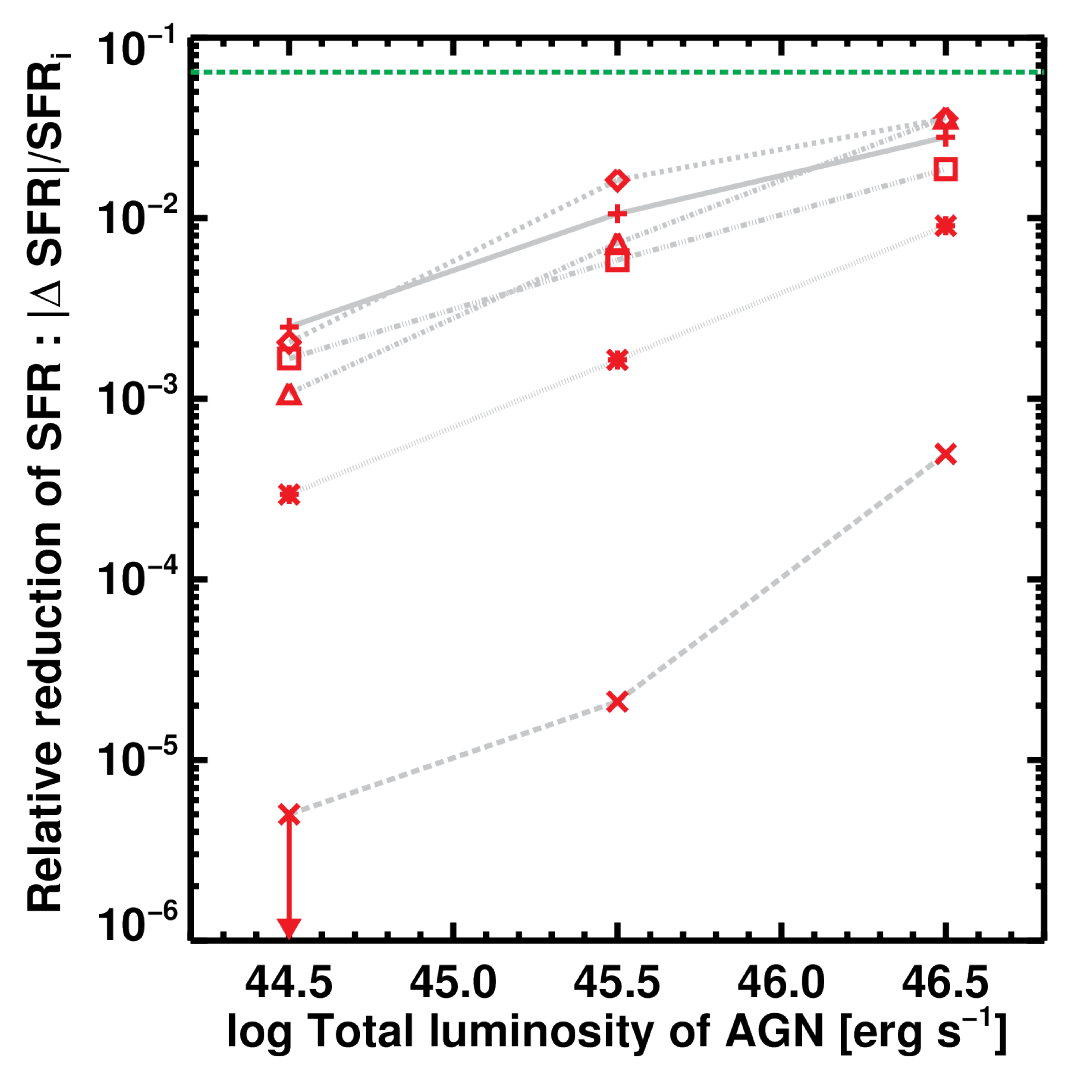}   
  \caption{{\bf Left:} SFR of the galaxy as a function of time. \textit{Green:} SFR with thermal AGN feedback, before RT. \textit{Red:} SFR with thermal AGN feedback, after RT, with AGN luminosity as labelled ($L_{AGN}=10^{44.5}$~erg~s$^{-1}$). \textit{Blue:} SFR without AGN feedback. The difference between the green and the blue curves shows the SFR fluctuations between two runs of a simulation$^\dagger$. Each symbol corresponds to one snapshot. For clarity, the error bars of the typical AGN (QSO) were shifted 1.5 Myrs to the left (right). Green error bars show the expected variability of the SFR. Red error bars account for the latter, plus resampling errors induced by the RT-process (see \citet{Roos2014} for further details). {\bf Right:} Relative reduction of  SFR as a function of AGN luminosity.  The green line shows the expected SFR variability between two runs of the same simulation. Clearly, the impact of AGN feedback (thermal+radiative) is not significant.  }
  \label{roos:fig2}
\end{figure}

Figure \ref{roos:fig2} (left) shows the temporal evolution of the SFR of the galaxy and confirms the idea that the bulk of the star-forming gas is left unaffected, since the SFR of the simulation with AGN feedback is only slightly reduced after the RT-process, for all snapshots studied, whatever the luminosity. Furthermore, this decrease is smaller than the typical variability of the SFR expected from such a simulation, as shown with the SFR of the run without AGN feedback.\footnote{Here, the SFR of the run with AGN feedback being higher than that without AGN feedback is unlikely due to SF-triggering and is likely caused by a random realisation of the cloud distribution. Such variations can be of a few percent of the total SFR.} Figure \ref{roos:fig2} (right), showing the relative reduction of SFR $\left|\Delta{SFR_{pre-post}}\right|/SFR_{pre-RT}$ as a function of AGN luminosity, further illustrates this behaviour. Indeed, it is clear that, even though there is an increasing trend with AGN luminosity, the reduction of the SFR due to both AGN photo-ionization and AGN heating is smaller than the expected variability of the SFR, and is therefore not significant.

\section{Impact of AGN feedback on 100 - 200 Myrs time scale star formation}

Despite the instantaneous character of our study, we can give clues about the SFR evolution on a time scale of a few 100 Myrs, by focusing on the impact of AGN radiation on the future sites of SF: reservoirs of atomic gas in the ISM, and in the envelopes of GMCs ($n = 0.3 - 10$~cm$^{-3}$; \citet{Dobbs2008a}). Such regions are likely to collapse on a few 100 Myr-scale and create new stars, only if there is no source of external heating, such as an AGN. The instantaneous impact of AGN radiation on atomic gas reservoirs is displayed in Table \ref{roos:table1}.  With a typical 1/3 AGN duty cycle, according to which the AGN is ``on''  with a typical AGN or strong AGN luminosity  1/3 of the time -- rare QSO episodes may also occur, it is very unlikely that the AGN will be able to quench SF on a time-scale of a few 100 Myrs. Nonetheless, if a long-lasting QSO episode occured ($\sim100$ Myrs, because of, e.g., a merger), the effects on future 100-Myr scale SF could be of greater importance.

\begin{table}[htp]
\footnotesize
\begin{minipage} [b]{.5\linewidth}
\caption{Effect of AGN on future (100~-~200 Myrs) star formation.  Rates are given for atomic gas ($0.3-10$ cm$^{-3}$).\label{roos:table1}}
\end{minipage}%
\begin{minipage} [b]{.5\linewidth}
\center
\begin{tabular}{|c|c|c|}
\hline
\footnotesize Regime & Heated mass rate & Ionized mass rate\\
\hline
\footnotesize Typical AGN &  $0-4$ \% & $0-2$ \%\\
\footnotesize Strong AGN &  $0.2-9$ \% & $0.01-3$ \%\\
\footnotesize Typical QSO & $2-30$ \% & $0.1-8$ \%\\
\hline
\end{tabular}
\hfill
\end{minipage}%
\end{table}

The impact on longer-term SF (up to 1 Gyr) depends on the ability of the AGN to keep the halo hot over an extended period of time, which can prevent gas supplies. As the halo of our simulated disk is not designed to be realistic, and the galaxy is not in its cosmological context, we cannot draw any conclusion about this topic.

\section{Conclusions}

We performed a complete treatment of the RT on 6 snapshots of a simulated SFG at high-redshift in post-processing. The high resolution ($\sim6$ pc) of the simulation allows us to probe the different phases of the ISM, from the diffuse halo, to the reservoirs of atomic gas and the GMCs.
Our main findings are as follows:
\begin{itemize}
\item The clumpy distribution of gas in the ISM plays a major role in the propagation of AGN radiation: while dense star-forming clumps can block AGN radiation at a very small scale length, diffuse interclump regions allow it to go further in the disk (up to 8 kpc in the QSO regime).
\item Most of the gas affected by the AGN (in the form of winds, heating, or photo-ionization) is diffuse, and located in the halo or the interclump medium. GMCs are  marginally heated, and thus the SFR reduction induced by both AGN heating and photo-ionization is of a few percent at most in the QSO regime. 
\item Moreover, even though the AGN generates powerful winds, no SF quenching is expected on a time scale of a few 100 Myrs under the assumption of a typical AGN duty cycle, since the AGN has a weak impact on the sites of future star formation (atomic gas).
\end{itemize}• 

\bibliographystyle{aa}  
\bibliography{roos_library} 

\begin{thebibliography}{20}
\expandafter\ifx\csname natexlab\endcsname\relax\def\natexlab#1{#1}\fi

\bibitem[{Booth \& Schaye(2009)}]{Booth2009}
Booth, C.~M. \& Schaye, J. 2009, Mon. Not. R. Astron. Soc., 398, 53

\bibitem[{{Di Matteo} {et~al.}(2005){Di Matteo}, Springel, \&
  Hernquist}]{DiMatteo2005}
{Di Matteo}, T., Springel, V., \& Hernquist, L. 2005, Nature, 1

\bibitem[{Dobbs \& Bonnell(2008)}]{Dobbs2008a}
Dobbs, C.~L. \& Bonnell, I.~a. 2008, Mon. Not. R. Astron. Soc., 385, 1893

\bibitem[{Dugan {et~al.}(2014)Dugan, Bryan, Gaibler, Silk, \& Haas}]{Dugan2014}
Dugan, Z., Bryan, S., Gaibler, V., Silk, J., \& Haas, M. 2014, Submitted to ApJ

\bibitem[{Elbaz {et~al.}(2009)Elbaz, Jahnke, Pantin, {Le Borgne}, \&
  Letawe}]{Elbaz2009}
Elbaz, D., Jahnke, K., Pantin, E., {Le Borgne}, D., \& Letawe, G. 2009, Astron.
  Astrophys., 1374, 1359

\bibitem[{Feain {et~al.}(2007)Feain, Papadopoulos, Ekers, \&
  Middelberg}]{Feain2007}
Feain, I.~J., Papadopoulos, P.~P., Ekers, R.~D., \& Middelberg, E. 2007,
  Astrophys. J., 872

\bibitem[{Ferland {et~al.}(2013)Ferland, Porter, van Hoof, Williams, Abel,
  Lykins, Shaw, Henney, \& Stancil}]{Ferland2013}
Ferland, G.~J., Porter, R.~L., van Hoof, P. A.~M., {et~al.} 2013, Rev. Mex.
  Astron. y Astrof\'{\i}sica, 49, 137

\bibitem[{Gabor \& Bournaud(2013)}]{Gabor2013b}
Gabor, J.~M. \& Bournaud, F. 2013, Mon. Not. R. Astron. Soc., 16, 1

\bibitem[{Gabor \& Bournaud(2014)}]{Gabor2014}
Gabor, J.~M. \& Bournaud, F. 2014, Mon. Not. R. Astron. Soc., 13, 1

\bibitem[{Gaibler {et~al.}(2012)Gaibler, Khochfar, Krause, \&
  Silk}]{Gaibler2012}
Gaibler, V., Khochfar, S., Krause, M., \& Silk, J. 2012, Mon. Not. R. Astron.
  Soc., 425, 438

\bibitem[{Hickox {et~al.}(2014)Hickox, Mullaney, Alexander, Chen, Civano,
  Goulding, Hainline, Alexander, Chen, Civano, Goulding, \&
  Hainline}]{Hickox2014}
Hickox, R.~C., Mullaney, J.~R., Alexander, D.~M., {et~al.} 2014, Astrophys. J.,
  782, 11

\bibitem[{Juneau {et~al.}(2013)Juneau, Dickinson, Bournaud, Alexander, Daddi,
  Mullaney, Magnelli, Kartaltepe, Hwang, Willner, Coil, Rosario, Trump, Weiner,
  Willmer, Cooper, Elbaz, Faber, Frayer, Kocevski, Laird, Monkiewicz, Nandra,
  Newman, Salim, \& Symeonidis}]{Juneau2013}
Juneau, S., Dickinson, M., Bournaud, F., {et~al.} 2013, Astrophys. J., 764, 176

\bibitem[{Maloney(1999)}]{Maloney1999}
Maloney, P.~R. 1999, Astrophys. Space Sci., 266, 207

\bibitem[{Mullaney {et~al.}(2012)Mullaney, Daddi, B\'{e}thermin, Elbaz, Juneau,
  Pannella, Sargent, Alexander, \& Hickox}]{Mullaney2012b}
Mullaney, J.~R., Daddi, E., B\'{e}thermin, M., {et~al.} 2012, Astrophys. J.,
  753, L30

\bibitem[{Proga {et~al.}(2014)Proga, Jiang, Davis, Stone, \& Smith}]{Proga2014}
Proga, D., Jiang, Y.-F., Davis, S.~W., Stone, J.~M., \& Smith, D. 2014,
  Astrophys. J., 780, 51

\bibitem[{Roos {et~al.}(2014)Roos, Juneau, Bournaud, \& Gabor}]{Roos2014}
Roos, O., Juneau, S., Bournaud, F., \& Gabor, J.~M. 2014, Submitted to ApJ,
  arXiv:1405.7971

\bibitem[{Rosdahl {et~al.}(2013)Rosdahl, Blaizot, Aubert, Stranex, \&
  Teyssier}]{Rosdahl2013}
Rosdahl, J., Blaizot, J., Aubert, D., Stranex, T., \& Teyssier, R. 2013, Mon.
  Not. R. Astron. Soc., 436, 2188

\bibitem[{Schawinski {et~al.}(2007)Schawinski, Thomas, Sarzi, Maraston,
  Kaviraj, Joo, Yi, \& Silk}]{Schawinski2007}
Schawinski, K., Thomas, D., Sarzi, M., {et~al.} 2007, Mon. Not. R. Astron.
  Soc., 18, 1

\bibitem[{Silk(2013)}]{Silk2013}
Silk, J. 2013, Astrophys. J., 772, 112

\bibitem[{Vogelsberger {et~al.}(2014)Vogelsberger, Genel, Springel, Torrey,
  Sijacki, Xu, Snyder, Nelson, \& Hernquist}]{Vogelsberger2014c}
Vogelsberger, M., Genel, S., Springel, V., {et~al.} 2014, Mon. Not. R. Astron.
  Soc., 444, 1518

\end{thebibliography}

\end{document}